\documentclass{revtex4}
\usepackage{amssymb}
\usepackage{amsfonts}
\usepackage{amsmath}

\input{tcilatex}

\begin{document}

\title{Klein-Gordon particles in G\"{o}del-type Som-Raychaudhuri cosmic
string spacetime and the phenomenon of spacetime associated degeneracies}
\author{Omar Mustafa}
\email{omar.mustafa@emu.edu.tr}
\affiliation{Department of Physics, Eastern Mediterranean University, G. Magusa, north
Cyprus, Mersin 10 - Turkey.}

\begin{abstract}
\textbf{Abstract:}\ We argue that only exact, comprehensive, and explicit
solutions for the fundamental models, the Klein-Gordon (KG) oscillators and
the KG-Coulomb, would help to understand the effects of gravitational fields
on the dynamics of such quantum mechanical systems. In the current
methodical proposal, the effects of the gravitational fields generated by a G%
\"{o}del-type Som-Raychaudhuri (SR) cosmic string spacetime on
KG-oscillators (KG-particles in general) are studied and reported. In so
doing, we revisit the KG-oscillators in a topologically trivial G\"{o}%
del-type spacetime background and use textbook procedures to report its
exact solution that covers even and odd parities. Next, we discuss the
drawbacks associated with the power series expansion approach that implies
the biconfluent Heun functions/polynomials solution. We, therefore,
recollect the so called pseudo perturbative shifted $\ell $ expansion
technique (PSLET) as an alternative and more sophisticated method/technique.
Illustrative examples are used: (i) a KG-oscillator in a topologically
trivial G\"{o}del-type spacetime, (ii) a quasi-free KG-oscillator in G\"{o}%
del SR-type cosmic string spacetime, (iii) a KG-Coulombic particle in G\"{o}%
del SR-type cosmic string spacetime at zero vorticity, and (iv) a massless
KG-particle in G\"{o}del SR-type cosmic string spacetime in a Cornell-type
Lorentz scalar potential. The corresponding exact energies are obtained from
the zeroth (leading) order correction of PSLET, where all higher order
correction identically vanish. The comprehensive exactness of the reported
solutions manifestly suggest degeneracies associated with spacetime (STAD)
phenomenon.

.


\end{abstract}

\maketitle

\section{Introduction}

Topological defects in spacetime have stimulated intensive research studies
in quantum gravity. Amongst are, domain walls \cite{Vilenkin 1983,Vilenkin
1985}, cosmic string \cite{Kibble 1976,Hiscock 1985}, global monopole \cite%
{Barriola 1989} and textures \cite{Cruz 2007}. The energy levels of
relativistic/non-relativistic quantum particles are shown to be affected by
the gravitational fields generated by different spacetime backgrounds with
such topological defects, not only in general relativity but also in the
geometrical theory of defects in condensed matter physics. In general
relativity, for example, the G\"{o}del spacetime metric \cite{Godel 1949},
with an embedded cosmic string, introduces itself as the first cosmological
solution to Einstein's equation with rotating matter. Its compact form
allowed analytical research studies of many physical and mathematical
systems in gravitational backgrounds with rotation and causality violation.
It has been shown that all spacetime homogeneous (ST-homogeneous) G\"{o}%
del-type metrics characterized by the vorticity $\Omega $ of spacetime (a
real number that represents uniform angular velocity or rotational material,
i.e, $\Omega =\pm \left\vert \Omega \right\vert $ \cite{R1}) and a given
value of the parameter $\tilde{\mu};-\infty \leq \tilde{\mu}^{2}\leq \infty $%
, can be transformed in cylindrical coordinates \cite{R1,Som 1968,Tiomno
1983,Drukker 2004,Carvalho 2014,Tiomno 1992,Wang 2015,Ahmed 2018,Inzunza
2022,Montigny 2018,Sedaghatnia 2019} to%
\begin{equation}
ds^{2}=-\left( dt+\alpha \Omega \frac{\sinh ^{2}\left( \tilde{\mu}\,r\right) 
}{\tilde{\mu}^{2}}d\varphi \right) ^{2}+\alpha ^{2}\frac{\sinh ^{2}\left( 2\,%
\tilde{\mu}\,r\right) }{4\,\tilde{\mu}^{2}}d\varphi ^{2}+dr^{2}+dz^{2}.
\label{Godel metric}
\end{equation}%
Where, for $\tilde{\mu}^{2}<0$ there is an infinite number of successive
causal and noncausal regions, and for $0\leq \tilde{\mu}^{2}<\Omega ^{2}$
there is one noncausal region for a given $r>r_{c}$. Here,  $r_{c}$ is the
critical radius determined by $g_{\varphi \varphi }=0$ (e.g.,  \cite{R1}) to
imply  
\begin{equation*}
\tanh \left( \tilde{\mu}\,r_{c}\right) =\frac{\tilde{\mu}\,}{\left\vert
\Omega \right\vert }\Longrightarrow r_{c}=\frac{1}{\tilde{\mu}\,}\text{%
arctanh}\left( \frac{\tilde{\mu}\,}{\left\vert \Omega \right\vert }\right) .
\end{equation*}%
For $\tilde{\mu}^{2}=\Omega ^{2}$ the G\"{o}del spacetime metric (the first
exact G\"{o}del solution of the Einstein equation describing a complete
causal, ST-homogeneous rotating universe \cite{Tiomno 1983}) is recovered,
and for $\tilde{\mu}<\left\vert \Omega \right\vert $ there is a causal
region (classified as chronologically safe) with radius $r_{c}$ surrounded
by the outer noncausal space  \cite{R1}  Nevertheless, at the limit $\tilde{%
\mu}\longrightarrow 0$ of the G\"{o}del spacetime metric (\ref{Godel metric}%
) we obtain the ST-homogeneous Som-Raychaudhuri (SR) solution%
\begin{equation}
ds^{2}=-\left( dt+\alpha \,\Omega \,r^{2}d\varphi \right) ^{2}+\alpha
^{2}\,r^{2}d\varphi ^{2}+dr^{2}+dz^{2}  \label{Som metric}
\end{equation}%
of the Einstein field equations \cite{Som 1968,R3,R4,R5}, in the presence of
cosmic string.\ Here, the disclination parameter $\alpha $ (curvature
parameter) admits the values $0<\alpha <1$ in general relativity (for cosmic
strings with positive curvature), $\alpha >1$ in the geometric theory of
defects in condensed matter (for a negative curvature), and $\Omega =0$ and $%
\alpha =1$ corresponds to Minkowski flat spacetime metric. Moreover, the
covariant and contravariant metric tensors associated with the
Som-Raychaudhuri spacetime are, respectively, given by%
\begin{equation}
g_{\mu \nu }=\left( 
\begin{tabular}{cccc}
$-1\smallskip $ & $\,0\,$ & $-\alpha \Omega r^{2}$ & $\,0$ \\ 
$0$ & $1\smallskip $ & $0$ & $0$ \\ 
$-\alpha \Omega r^{2}$ & $\,0$ & $\,\alpha ^{2}r^{2}\left( 1-\Omega
^{2}r^{2}\right) \,$ & $0$ \\ 
$0$ & $0$ & $0$ & $1$%
\end{tabular}%
\right) \Longleftrightarrow g^{\mu \nu }=\left( 
\begin{tabular}{cccc}
$\left( \Omega ^{2}r^{2}-1\smallskip \right) $ & $\,0\,$ & $-\frac{\Omega }{%
\alpha }$ & $0$ \\ 
$0$ & $\,1\smallskip $ & $0$ & $0$ \\ 
$-\frac{\Omega }{\alpha }$ & $\,0$ & $\,\frac{1}{\alpha ^{2}r^{2}}\,$ & $0$
\\ 
$0$ & $0$ & $0\,$ & $\,1$%
\end{tabular}%
\right) ;\ \det \left( g\right) =-\alpha ^{2}r^{2}.  \label{som tensor}
\end{equation}

The objective of the current methodical proposal is to explore in a
comprehensive manner the effect of the gravitational field generated by\ the
G\"{o}del-type cosmic string SR-spacetime (\ref{Som metric}) on the spectra
of some Klein-Gordon (KG) particles of fundamental quantum gravity nature,
like the KG-oscillators and KG-Coulomb models. Such fundamental models have
been studied by different authors (e.g., \cite{Carvalho 2014,Tiomno
1992,Wang 2015,Ahmed 2018,Inzunza 2022,Montigny 2018,Sedaghatnia
2019,Vitoria 2018,Neto 2020,Ahmed 2019}) where the vorticity parameter $%
\Omega $ is assumed to take the form $\Omega \geq 0$ and the energies are
assumed to be positive $E=+\left\vert E\right\vert $ or $E=\pm \left\vert
E\right\vert $ where all nodeless states are lost in the process (excluding
Hassanabadi et al \cite{Montigny 2018,Sedaghatnia 2019} who have included
all nodeless states with the radial quantum number $n_{r}=0$). However, in
Duarte de Oliveira et al \cite{R1}, $\Omega $ is a real number that denotes
uniform angular velocity or rotation of material (i.e., $\Omega =\pm
\left\vert \Omega \right\vert $). In the current methodical proposal, we
discuss KG-particles' spectral properties under the influence of the
gravitational field generated by\ the G\"{o}del-type cosmic string
SR-spacetime (\ref{Som metric}) taking into account $E=\pm \left\vert
E\right\vert $, $\Omega =\pm \left\vert \Omega \right\vert $, and including
nodeless states. In so doing, nevertheless, it is unavoidably inviting to
reconsider and report on the KG-oscillators in a topologically trivial G\"{o}%
del-type spacetime metric, used by Ahmed \cite{Ahmed 2018},%
\begin{equation}
ds^{2}=-dt^{2}+dx^{2}+\left( 1-\alpha ^{2}x^{2}\right) dy^{2}-2\alpha
xdtdy+dz^{2},  \label{Ahmed metric1}
\end{equation}%
where $H\left( x\right) =\alpha x$ and $D\left( x\right) =1$ in the general\
G\"{o}del metric \cite{Godel 1949,Tiomno 1983}%
\begin{equation*}
ds^{2}=-\left( dt+H\left( x\right) dy\right) ^{2}+dx^{2}+D\left( x\right)
^{2}dy^{2}+dz^{2},
\end{equation*}%
using the coordinates $\left( x^{0}=t,x^{1}=x,x^{2}=y,x^{3}=z\,;\;x,y,z\in
\left( -\infty ,\infty \right) \right) $. Where, the covariant and
contravariant elements of the spacetime metric tensors are given by%
\begin{eqnarray}
g_{00} &=&-1,\,g_{11}=g_{33}=1,\,g_{02}=g^{20}=-\alpha x,\,g_{22}=1-\alpha
^{2}x^{2},  \notag \\
\,g^{00} &=&\alpha ^{2}x^{2}-1,\,g^{11}=g^{22}=g^{33}=1,\text{ }\det \left(
g\right) =-1.  \label{1D metric elements}
\end{eqnarray}

The organization of our methodical proposal is in order. We start with the
KG-oscillators in a topologically trivial G\"{o}del-type spacetime (\ref%
{Ahmed metric1}) considered by Ahmed \cite{Ahmed 2018}, who has reported
only the positive energy states and missed not only the ground state but
also the negative energy states. For the sake of scientific correctness, we
give this problem a proper textbook quantum mechanical treatment \cite%
{Flugge 1974} and report general forms of the spectra, in section 2, that
include all positive and negative energy states. In section 3, we discuss
KG-particles in G\"{o}del SR-type cosmic string spacetime in a Cornnel-type
Lorentz scalar potentials and show the disadvantages of using a power series
(Frobinous method) expansion method that leads to biconfluent Heun
functions/polynomials. In section 4, we recollect, as a powerful alternative
method, the so called pseudo-perturbative shifted-$\ell $ expansion
technique (PSLET), to solve for Schr\"{o}dinger equation (a common form that
relativistic wave equations collapse into) in $D$-dimensions \cite%
{Herschbach 1993,Imbo 1983,Imbo 1984,Mustafa 1998,Mustafa 2000,Mustafa
1993,Mustafa 1994,Mustafa 1991} to serves our current methodical proposal.
PSLET has shown, with a brut force evidence through comparison with
numerical integration techniques, to provide very highly accurate results
(even with its non-Hermitian $\mathcal{PT}$-symmetric version by Mustafa and
Znojil \cite{Mustafa 2002,Znojil 2002}). We use some illustrative examples,
in section 5, that include a KG-oscillator (of section 2) in a topologically
trivial G\"{o}del-type spacetime (\ref{Ahmed metric1}), a quasi-free
KG-oscillator in G\"{o}del SR-type cosmic string spacetime (\ref{Som metric}%
), a KG-Coulombic particle in G\"{o}del SR-type cosmic string spacetime (\ref%
{Som metric}) at zero vorticity (i.e., $\Omega =0$), and a massless
KG-particles in G\"{o}del SR-type cosmic string spacetime in a Cornell-type
Lorentz scalar potential \cite{Quigg 1979}. We conclude in section 6.

\section{KG-oscillators in a topologically trivial G\"{o}del-type spacetime
background}

\subsection{KG-oscillator I as a topologically trivial G\"{o}del-type
spacetime byproduct:}

In this section we recollect the topologically trivial G\"{o}del-type
spacetime metric (\ref{Ahmed metric1}), along with the metric tensor
elements in (\ref{1D metric elements}) and $\det \left( g\right) =-1$, used
by Ahmed \cite{Ahmed 2018}. Then, a KG-particle of rest mass energy $m$
(denoting $mc^{2}$ in $c=\hbar =1$ units) in such a G\"{o}del-type spacetime
is described by the KG-equation%
\begin{equation}
\frac{1}{\sqrt{-g}}\partial _{\mu }\left( \sqrt{-g}g^{\mu \nu }\partial
_{\nu }\Psi \right) =m^{2}\Psi .  \label{KG-eq}
\end{equation}%
Consequently, with 
\begin{equation}
\Psi \left( t,x,y,z\right) =\exp \left( i\left[ k_{y}y+k_{z}z-Et\right]
\right) \phi \left( x\right) ,  \label{psi(x)}
\end{equation}%
it yields%
\begin{equation}
\phi ^{\prime \prime }\left( x\right) -\left( \gamma ^{2}x^{2}+2\,\gamma
k_{y}\,x\right) \,\phi \left( x\right) =\grave{\beta}\,\phi \left( x\right)
;\,\,\grave{\beta}=m^{2}+k_{y}^{2}+k_{z}^{2}-E^{2},\;\gamma ^{2}=\alpha
^{2}E^{2}.  \label{1D-KG-oscillator}
\end{equation}%
At this point, one should notice that $\gamma =\pm \alpha \left\vert
E\right\vert $ as mandated by the very nature of the G\"{o}del-type
spacetime in (\ref{Ahmed metric1}). Obviously, this equation resembles the
well known one-dimensional Schr\"{o}dinger equation for the so called
shifted-oscillator (but does not inherit its parametric characterizations)
that can be treated, in a straightforward textbook manner, by rewriting (\ref%
{1D-KG-oscillator}) as 
\begin{equation}
\phi ^{\prime \prime }\left( x\right) -\left( \gamma x+k_{y}\right)
^{2}\,\phi \left( x\right) =\tilde{\beta}\,\phi \left( x\right) \,;\,\,%
\tilde{\beta}=\grave{\beta}-k_{y}^{2}=m^{2}+k_{z}^{2}-E^{2}.
\label{KG-shifted oscillator}
\end{equation}%
One would then use the change of variable $u=\gamma x+k_{y}$ to obtain%
\begin{equation}
\phi ^{\prime \prime }\left( u\right) -\tilde{\gamma}^{2}\,u^{2}\,\phi
\left( u\right) +\lambda \,\phi \left( u\right) =0\,;\,\,\lambda =-\frac{%
\tilde{\beta}}{\gamma ^{2}}\,,\tilde{\gamma}^{2}=\frac{1}{\gamma ^{2}}\,.
\label{KG-oscillator u}
\end{equation}%
This is the one-dimensional Schr\"{o}dinger-like harmonic oscillator form
that admits exact textbook solution with the eigenvalues and eigenfunctions,
respectively, given by 
\begin{equation}
\lambda _{n}=\left\vert \tilde{\gamma}\right\vert \left( 2n+1\right) ,\text{
and}\;\phi \left( u\right) =\mathcal{N}_{n}\,e^{-\frac{\left\vert \tilde{%
\gamma}\right\vert \,u^{2}}{2}}\,H_{n}\left( \sqrt{\left\vert \tilde{\gamma}%
\right\vert }\,u\right) ;\;n=0,1,2,\cdots ,  \label{HO-solution}
\end{equation}%
where $H_{n}\left( \sqrt{\left\vert \tilde{\gamma}\right\vert }\,u\right) $
are the Hermite polynomials and $\mathcal{N}_{n}$ are the normalization
constants. The finiteness and square integrability of the wave function
enforces the quantum mechanical condition that $\left\vert \tilde{\gamma}%
\right\vert =\pm \tilde{\gamma}_{\pm }\geq 0$ (i.e., $\tilde{\gamma}=\tilde{%
\gamma}_{\pm };\;\tilde{\gamma}_{+}=1/\alpha E_{+}$, $\tilde{\gamma}%
_{-}=1/\alpha E_{-}\Longrightarrow \left\vert \tilde{\gamma}\right\vert =\pm 
\tilde{\gamma}_{\pm }$, where $E_{\pm }=\pm \left\vert E\right\vert $).
Consequently, with the parametric substitutions above (\ref{1D-KG-oscillator}%
), one obtains%
\begin{equation}
\frac{m^{2}+k_{z}^{2}-E_{\pm }^{2}}{\alpha ^{2}E_{\pm }^{2}}=-\left\vert 
\tilde{\gamma}\right\vert \left( 2n+1\right) \Rightarrow E_{\pm }^{2}\mp
\alpha E_{\pm }\left( 2n+1\right) -\left( m^{2}+k_{z}^{2}\right)
=0;\;n=0,1,2,\cdots .  \label{Exact KG-HO-energies 1}
\end{equation}%
Hence,%
\begin{equation}
E_{n}=E_{\pm }=\pm \alpha \left( n+\frac{1}{2}\right) \pm \sqrt{\alpha
^{2}\left( n+\frac{1}{2}\right) ^{2}+m^{2}+k_{z}^{2}}.
\label{Exact KG-HO-energies}
\end{equation}%
Which in turn implies%
\begin{equation}
E_{+}=+\alpha \left( n+\frac{1}{2}\right) +\sqrt{\alpha ^{2}\left( n+\frac{1%
}{2}\right) ^{2}+m^{2}+k_{z}^{2}}  \label{Sec2a1}
\end{equation}%
as the positive energy solution, and%
\begin{equation}
E_{-}=-\alpha \left( n+\frac{1}{2}\right) -\sqrt{\alpha ^{2}\left( n+\frac{1%
}{2}\right) ^{2}+m^{2}+k_{z}^{2}}  \label{Sec2a2}
\end{equation}%
for the negative energy solution. Notably, unlike the solution reported by
Ahmed \cite{Ahmed 2018} (see equation (58) in \cite{Ahmed 2018}), the
textbook solution above includes both even and odd parities, positive and
negative energies, as well as the ground state \cite{Flugge 1974}. 

\subsection{KG-oscillator II as a topologically trivial G\"{o}del-type
spacetime and momentum operator deformation byproduct}

Using similar recipe as that of Moshinsky and Szczepaniak \cite{Moshinsky
1989} and Mirza and Mohadesi \cite{Mirza 2004} for the Dirac and KG
oscillators, we redefine the momentum operator as%
\begin{equation}
p_{\mu }\longrightarrow p_{\mu }+i\eta \chi _{\mu };\;\chi _{\mu }=\left(
0,x,0,0\right) ,  \label{HO-momentum}
\end{equation}%
where\ the dimensions of the second term suits the dimensions of the
momentum operator $p_{\mu }$ and here we consider $\eta $ to have positive
and/or negative values. Then the KG-equation reads%
\begin{equation}
\frac{1}{\sqrt{-g}}\left( \partial _{\mu }+\eta \chi _{\mu }\right) \sqrt{-g}%
g^{\mu \nu }\left( \partial _{\nu }-\eta \chi _{\nu }\right) \Psi =m^{2}\Psi
.  \label{KG-oscillator2}
\end{equation}%
At this point, one may use $\eta =m\omega \geq 0$ (with $m$ denoting rest
mass of the KG-particle) to recover the traditionally used values as in
(e.g., \cite{Ahmed 2018,Vitoria 2018,Neto 2020,Ahmed 2019,Moshinsky
1989,Mirza 2004} and\ other related references cited therein). However, we
shall use a general parameter $\eta $ and allow it to take positive and/or
negative values. Yet, in this case we avoid eminent confusion and
inconsistency between $m^{2}$ (denoting $m^{2}c^{2}=mmc^{2}$, rest mass
multiplied by rest mass energy) on the R.H.S. and the rest mass of the
particle on the L.H.S. of the KG-equation (\ref{KG-oscillator2}) for $\eta
=m\omega $. This point is made clear by Moshinsky and Szczepaniak  \cite%
{Moshinsky 1989} and Mirza and Mohadesi \cite{Mirza 2004} while dealing with
the Dirac and KG oscillators. We therefore stick with our assumption and use
the spacetime metric tensor elements in (\ref{1D metric elements}), to
recast (\ref{KG-oscillator2}) as 
\begin{equation}
\phi ^{\prime \prime }\left( x\right) -\left( \tilde{\omega}%
^{2}x^{2}+2\,\gamma k_{y}\,x\right) \,\phi \left( x\right) =\zeta \,\phi
\left( x\right) ;\,\,\zeta =m^{2}+k_{y}^{2}+k_{z}^{2}-E^{2}+\eta ,\;\tilde{%
\omega}^{2}=\gamma ^{2}+\eta ^{2},\;\gamma =\alpha E.  \label{KG-oscillator3}
\end{equation}%
Hereby, one should be aware that $\eta ^{2}x^{2}$ would take the form of $%
\eta ^{2}x^{2}=m\left( m\omega ^{2}x^{2}\right) $ as in \cite{Moshinsky
1989,Mirza 2004} when the speed of light is kept intact in Dirac and/or KG
equations (i.e., consistent with $m^{2}c^{2}=m\left( mc^{2}\right) $ on the
R.H.S. of (\ref{KG-oscillator3})). The source of confusion is therefore
clear and should be avoided. Moreover, equation (\ref{KG-oscillator3})
admits solution in the form of hypergeometric functions \cite{Flugge 1974}
given by%
\begin{equation}
\,\phi \left( x\right) =e^{-\frac{\left\vert \tilde{\omega}\right\vert x^{2}%
}{2}-\frac{\delta x}{\left\vert \tilde{\omega}\right\vert }}\left[
A\,\,_{1}F_{1}\left( \tilde{a},\frac{1}{2},\frac{\left( \tilde{\omega}%
^{2}x+\delta \right) ^{2}}{\left\vert \tilde{\omega}\right\vert ^{3}}\right)
+B\,\,\left( \tilde{\omega}^{2}x+\delta \right) \,_{1}F_{1}\left( \tilde{b},%
\frac{3}{2},\frac{\left( \tilde{\omega}^{2}x+\delta \right) ^{2}}{\left\vert 
\tilde{\omega}\right\vert ^{3}}\right) \right] ,  \label{KG psi 3}
\end{equation}%
where $\delta =\gamma k_{y}=\alpha Ek_{y}$, $\left\vert \tilde{\omega}%
\right\vert \geq 0$ to secure finiteness and square integrability of the
wave function, and%
\begin{equation}
\tilde{b}=\frac{3}{4}-\frac{\delta ^{2}}{4\left\vert \tilde{\omega}%
\right\vert ^{3}}+\frac{\zeta }{4\left\vert \tilde{\omega}\right\vert }=%
\tilde{a}+\frac{1}{2}.
\end{equation}%
Again, the condition for the hypergeometric functions to become polynomials
of degree $n\geq 0$ suggests that $a=-n$ and $b=a+1/2=-n$. This would imply
that%
\begin{equation}
\tilde{a}=\frac{1}{4}-\frac{\delta ^{2}}{4\left\vert \tilde{\omega}%
\right\vert ^{3}}+\frac{\zeta }{4\left\vert \tilde{\omega}\right\vert }%
=-n\Longrightarrow \zeta _{2n}=-\left\vert \tilde{\omega}\right\vert \left(
4n+1\right) +\frac{\delta ^{2}}{\tilde{\omega}^{2}},  \label{at}
\end{equation}%
for even parity, and%
\begin{equation}
\tilde{b}=-n\Longrightarrow \zeta _{2n+1}=-\left\vert \tilde{\omega}%
\right\vert \left( 4n+3\right) +\frac{\delta ^{2}}{\tilde{\omega}^{2}}.
\label{bt}
\end{equation}%
for odd parity solutions. The two results (\ref{at}) and (\ref{bt}) would
reduce to a common form 
\begin{equation}
\zeta _{n}=-\left\vert \tilde{\omega}\right\vert \,\left( 2n+1\right) +\frac{%
\delta ^{2}}{\tilde{\omega}^{2}};\;\left\vert \tilde{\omega}\right\vert =\pm
\alpha E_{\pm }\sqrt{1+\frac{\eta ^{2}}{\alpha ^{2}E_{\pm }^{2}}}\geq 0,
\end{equation}%
to imply that%
\begin{equation}
E_{\pm }^{2}-\left( m^{2}+k_{y}^{2}+k_{z}^{2}+\eta \right) =\pm \alpha
E_{\pm }\sqrt{1+\frac{\eta ^{2}}{\alpha ^{2}E_{\pm }^{2}}}\,\left(
2n+1\right) -\frac{\alpha ^{2}E_{\pm }^{2}k_{y}^{2}}{\alpha ^{2}E_{\pm
}^{2}+\eta ^{2}}.  \label{Energy eq}
\end{equation}
Obviously, this result collapses into that in (\ref{Exact KG-HO-energies})
for $\eta =0$, where the effect of even and odd parities is reflected on the
related spectrum. 

\section{KG-particles in a 4-vector and scalar Lorentz potentials in G\"{o}%
del SR-type cosmic string spacetime background}

The KG-equation for a spin-0 particle in a 4-vector $A_{\mu }$ and a radial
scalar $S\left( r\right) $ potentials in the G\"{o}del SR-type spacetime
background (\ref{Som metric})\ is given by%
\begin{equation}
\frac{1}{\sqrt{-g}}D_{\mu }\left( \sqrt{-g}g^{\mu \nu }D_{\nu }\Psi \right)
=\left( m+S(r)\right) ^{2}\Psi ,  \label{EM-KG eq}
\end{equation}%
where the gauge-covariant derivative is given by $D_{\mu }=\partial _{\mu
}-ieA_{\mu }$. This would in turn allow us to cast equation (\ref{EM-KG eq})
as%
\begin{equation}
\left\{ -D_{t}^{2}+\frac{1}{r}\,D_{r}\left( r\,D_{r}\right) +\left( \Omega
\,rD_{t}-\frac{1}{\alpha r}D_{\varphi }\right) ^{2}+D_{z}^{2}-\left(
m+S(r)\right) ^{2}\right\} \Psi =0.  \label{General-PDM-KG}
\end{equation}%
At this point, we may now define the corresponding gauge-covariant
derivatives so that%
\begin{equation}
\begin{tabular}{llll}
$D_{t}=\partial _{t}-ieA_{t}=\partial _{t}-iV\left( r\right) ,$ & $%
D_{r}=\partial _{r},$ & $D_{\varphi }=\partial _{\varphi }-ieA_{\varphi },$
& $D_{z}=\partial _{z},$%
\end{tabular}
\label{covariant derivatives}
\end{equation}%
where, $V\left( r\right) =eA_{t}$ is the Lorentz 4-vector potential (i.e.,
transforms within the 4-vector potential $A_{\mu }$), $S\left( r\right) $ is
the Lorentz scalar potential (i.e., transforms like the rest mass energy $%
m\longrightarrow m+S\left( r\right) $), and $eA_{\varphi }$ may include both
magnetic and Aharonov-Bohm flux fields effects \cite{Mustafa3 2021,Mustafa
Algadhi1 2020}. We now use the assumption that%
\begin{equation}
\Psi \left( t,r,\varphi ,z\right) =\exp \left( i\left[ \ell \varphi
+k_{z}z-Et\right] \right) \psi \left( r\right) =\exp \left( i\left[ \ell
\varphi +k_{z}z-Et\right] \right) \frac{R\left( r\right) }{\sqrt{r}}
\label{Psi(r) 1}
\end{equation}%
in (\ref{General-PDM-KG}) to obtain%
\begin{equation}
\left\{ \partial _{r}^{2}+\frac{1}{4r^{2}}+\left( E+V\left( r\right) \right)
^{2}-\left[ \Omega r\left( E+V\left( r\right) \right) +\frac{\left( \ell
-eA_{\varphi }\right) }{\alpha r}\right] ^{2}-k_{z}^{2}-\left( m+S(r)\right)
^{2}\right\} R\left( r\right) =0,  \label{PDM-KG1}
\end{equation}%
This result would describe KG-particles in G\"{o}del SR-type cosmic string
spacetime background.

In what follows, we shall be interested in a set of KG-particles in G\"{o}%
del SR-type cosmic string spacetime under the influence of $A_{\varphi }=0$
(i.e., no magnetic or Aharonov-Bohm flux fields effects), and no Lorentz
4-vector potential (i.e., $V\left( r\right) =0$). Under such settings, the
KG-equation (\ref{PDM-KG1}) reduces to 
\begin{equation}
\left\{ \partial _{r}^{2}-\frac{\left( \tilde{\ell}^{2}-1/4\right) }{r^{2}}%
-\Omega ^{2}E^{2}r^{2}-2mS\left( r\right) -S(r)^{2}+\tilde{\lambda}\right\}
R\left( r\right) =0;\;\tilde{\ell}=\frac{\ell }{\alpha },\,\tilde{\lambda}%
=E^{2}-2\,\tilde{\ell}\,\Omega \,E-k_{z}^{2}-m^{2}.  \label{massless KG}
\end{equation}%
Obviously, this equation resembles the one-dimensional form of the
two-dimensional radial Schr\"{o}dinger equation. Which, with a Cornell-type
Lorentz scalar potential $S\left( r\right) =ar+b/r$ would result%
\begin{equation}
\left\{ \partial _{r}^{2}-\frac{\left( \mathcal{L}^{2}-1/4\right) }{r^{2}}-%
\tilde{\Omega}^{2}r^{2}-2mar-\frac{2mb}{r}+\Lambda \right\} R\left( r\right)
=0,  \label{Heun1}
\end{equation}%
where%
\begin{equation}
\tilde{\Omega}^{2}=\Omega ^{2}E^{2}+a^{2},\;\mathcal{L}^{2}=\tilde{\ell}%
^{2}+b^{2},\,\Lambda =E^{2}-2\,\tilde{\ell}\,\Omega \,E-\left(
k_{z}^{2}+m^{2}+2ab\right) .  \label{Heun2}
\end{equation}%
Equation (\ref{Heun1}) is known to admit, using power series method, a
solution in the form of biconfluent Heun functions%
\begin{equation}
R\left( r\right) =.C\,r^{\left\vert \mathcal{L}\right\vert +1/2}\,\exp
\left( -\frac{\left\vert \tilde{\Omega}\right\vert }{2}r^{2}-\frac{am}{%
\left\vert \tilde{\Omega}\right\vert }r\right) \,H_{B}\left( \grave{\alpha},%
\grave{\beta},\grave{\gamma},\grave{\delta},\sqrt{\left\vert \tilde{\Omega}%
\right\vert }r\right) .  \label{Heun3}
\end{equation}%
where $\grave{\alpha}=2\left\vert \mathcal{L}\right\vert ,\,\grave{\beta}=%
\frac{2am}{\left\vert \tilde{\Omega}\right\vert ^{3/2}},\,\grave{\gamma}=%
\frac{a^{2}m^{2}+\Lambda \tilde{\Omega}^{2}}{\left\vert \tilde{\Omega}%
\right\vert ^{3}},$ and $\delta =\frac{4mb}{\sqrt{\left\vert \tilde{\Omega}%
\right\vert }}$. For the case $a=0=b$, this solution implies 
\begin{eqnarray}
R\left( r\right)  &=&.C\,r^{\left\vert \tilde{\ell}\right\vert +1/2}\,\exp
\left( -\frac{\left\vert \Omega E\right\vert }{2}r^{2}\right) \,H_{B}\left(
2\left\vert \tilde{\ell}\right\vert ,0,\frac{\Lambda }{\left\vert \Omega
E\right\vert },0,\sqrt{\left\vert \Omega E\right\vert }r\right)   \notag \\
&=&.C\,r^{\left\vert \tilde{\ell}\right\vert +1/2}\,\exp \left( -\frac{%
\left\vert \Omega E\right\vert }{2}r^{2}\right) \,_{1}F_{1}\left( \left[ 
\frac{1}{2}+\frac{\grave{\alpha}}{4}-\frac{\grave{\gamma}}{4}\right] ,\left[
1+\frac{\grave{\alpha}}{2}\right] ,\left\vert \Omega E\right\vert
r^{2}\right) ,  \label{Heun4}
\end{eqnarray}%
where the condition that 
\begin{equation}
\frac{1}{2}+\frac{\grave{\alpha}}{4}-\frac{\grave{\gamma}}{4}=\frac{1}{2}+%
\frac{\left\vert \tilde{\ell}\right\vert }{2}-\frac{\Lambda }{4\left\vert
\Omega E\right\vert }=-n_{r}  \label{Heun4-1}
\end{equation}%
is required for the confluent hypergeometric function to become a polynomial
of degree $n_{r}\geq 0$ and to secure the finiteness and square
integrability of the wave function. This would immediately result, in terms
of the associated Laguerre polynomials $\,L_{n_{r}}^{\left\vert \tilde{\ell}%
\right\vert }\left( z^{2}\right) $, that%
\begin{equation}
\Lambda =2\left\vert \Omega E\right\vert \left( 2n_{r}+\left\vert \tilde{\ell%
}\right\vert +1\right) ,\text{ and }R\left( r\right) =.C\,r^{\left\vert 
\tilde{\ell}\right\vert +1/2}\,\exp \left( -\frac{\left\vert \Omega
E\right\vert }{2}r^{2}\right) \,L_{n_{r}}^{\left\vert \tilde{\ell}%
\right\vert }\left( \left\vert \Omega E\right\vert r^{2}\right) 
\label{Heun5}
\end{equation}%
as the exact eigen energies and eigen functions for the two dimensional Schr%
\"{o}dinger harmonic oscillator. Moreover, the same condition (\ref{Heun4-1}%
) should also be imposed on the biconfluent Heun function (\ref{Heun3}) so
that a biconfluent Heun polynomial \cite{R2} of degree $n=2n_{r}\geq 0$ is
obtained (c.f., e.g., \cite{Neto 2020,Mustafa3 2021,Mustafa 2022,Mustafa2
2021}) to imply%
\begin{equation}
\Lambda =2\left\vert \Omega E\right\vert \,\sqrt{1+\frac{a^{2}}{\Omega
^{2}E^{2}}}\left( 2n_{r}+\left\vert \mathcal{L}\right\vert +1\right) ,
\label{Heun6-1}
\end{equation}%
and%
\begin{equation}
E^{2}-2\,\tilde{\ell}\,\Omega \,E-\left( k_{z}^{2}+m^{2}+2ab\right)
=2\left\vert \Omega E\right\vert \,\sqrt{1+\frac{a^{2}}{\Omega ^{2}E^{2}}}%
\left( 2n_{r}+\left\vert \mathcal{L}\right\vert +1\right) .  \label{Heun6}
\end{equation}
Where the detailed analysis of such a quadratic equation are given in
section 5-B, taking into account that%
\begin{equation}
\left\vert \Omega E\right\vert =\left\{ 
\begin{tabular}{l}
$+\Omega _{\pm }E_{\pm }\smallskip $ \\ 
$-\Omega _{\mp }E_{\pm }$%
\end{tabular}%
\right\} \geq 0;\;E_{\pm }=\pm \left\vert E\right\vert ,\;\Omega _{\pm }=\pm
\left\vert \Omega \right\vert ,  \label{Heun6+1}
\end{equation}
At this point, the truncation order $n$ of a power series (biconfluent Heun
function here) does not make it a valid quantum number, but rather it should
be correlated to the well known quantum numbers as done above (i.e., $%
n=2n_{r}\geq 0$). Hereby, one should give credentials for Neto and
co-workers \cite{Neto 2020} as they resemble a group of the very few who
have correctly used condition (\ref{Heun4-1}) on the biconfluent Heun
polynomials. Only under such condition that the result of (\ref{Heun6})
would recover those of (\ref{Heun5}) for $a=0=b$. Obviously, however, for
the case $a=0=\Omega $ the solution in (\ref{Heun6}) collapses into a free
relativistic particle eigenvalues 
\begin{equation}
E^{2}-\left( k_{z}^{2}+m^{2}\right) =0,  \label{Heun7}
\end{equation}%
although we have an effective two-dimensional Coulomb problem (i.e., a
KG-Coulombic particle). This is a clear drawback of the Heun power series
(Frobinous method) expansion approach. Moreover, if condition (\ref{Heun4-1}%
) is not implemented then the biconfluent Heun polynomial solution would
fail to address the exact solution for the KG-oscillator for $a=0=b$. One
should, therefore, search for an alternative method to deal with such a case
in (\ref{Heun1}).

As a more reliable alternative method, to the power series (Frobinous
method) expansion, we recollect the so called pseudo-perturbative shifted-$%
\ell $ expansion technique (PSLET) for the $D$-dimensional radial Schr\"{o}%
dinger-type equation and solve for different KG-particles' settings in the G%
\"{o}del SR-type cosmic string spacetime backgrounds of (\ref{Heun1}). In
order to make our proposal self contained, we, in short, discuss PSLET in
the following section. We shall also use Ahmed's model (\ref{KG-oscillator u}%
) \cite{Ahmed 2018} as one of the illustrative examples to be used in the
current methodical proposal.

\section{PSLET for the $D$-dimensional Schr\"{o}dinger-type equation}

Let us recollect that the $D$-dimensional radial Schr\"{o}dinger equation ($%
\hbar =2m=1$ units are used in this section) reads%
\begin{equation}
\left\{ -\partial _{r}^{2}+\frac{\ell _{D}\left( \ell _{D}+1\right) }{r^{2}}%
+V\left( r\right) \right\} \Phi _{k,\ell }\left( r\right) =E_{k,\ell }\Phi
_{k,\ell }\left( r\right) ,  \label{D-dimensional Schrodinger}
\end{equation}%
where $\ell _{D}=\ell +\left( D-3\right) /2$ to incorporate interdimensional
degeneracies associated with the isomorphism between angular momentum and
the dimensionality $D$. Hereby, one should notice that for $D=1$, $\ell
_{1}=\ell -1$ and hence $\ell _{D}\left( \ell _{D}+1\right) =\ell \left(
\ell -1\right) =0$ for $\ell =0$ and $\ell =1$ to denote even and odd
parity, respectively, for the one-dimensional case with $r=x\in \left(
-\infty ,\infty \right) $. In the two dimensional case with radially
cylindrical symmetry, $r=\sqrt{x^{2}+y^{2}}\in \left( 0,\infty \right) $ and 
$\ell _{2}=\left\vert \ell \right\vert -1/2$ with $\ell $ denoting the
magnetic quantum number. Moreover, in the three-dimensional radially
spherical symmetry, $r=\sqrt{x^{2}+y^{2}+z^{2}}\in \left( 0,\infty \right) $
and $\ell _{3}=\ell $ where $\ell $ denotes angular momentum quantum number.
Since we are going to use $1/\ell _{D}$ as a small perturbation expansion
parameter at large-$\ell _{D}$ limit (i.e., pseudo-classical limit, e.g., 
\cite{Imbo 1983,Imbo 1984,Herschbach 1993,Mustafa 1998,Mustafa 2000,Mustafa
2002,Mustafa 1993,Mustafa 1994,Mustafa 1991}) we avoid the trivial case for $%
\ell _{D}=0$ (for $D=3$ and $\ell =0$) and use a shift $\beta $ so that $%
\breve{\ell}=\ell _{D}-\beta $. Under such settings, our Schr\"{o}dinger
equation in (\ref{D-dimensional Schrodinger}) reads%
\begin{equation}
\left\{ -\partial _{r}^{2}+\frac{\breve{\ell}^{2}+\breve{\ell}\,\left(
2\beta +1\right) +\beta \left( \beta +1\right) }{r^{2}}+V\left( r\right)
\right\} \Phi _{k,\ell }\left( r\right) =E_{k,\ell }\Phi _{k,\ell }\left(
r\right) .  \label{P1}
\end{equation}%
For convenience, we now shift the origin of the coordinate system and use%
\begin{equation}
x=\frac{\breve{\ell}^{1/2}}{r_{\circ }}\left( r-r_{\circ }\right) ,
\label{P2}
\end{equation}%
and expand about $x=0$ to obtain%
\begin{equation}
\left( \frac{x}{\breve{\ell}^{1/2}}+1\right) ^{-2}=1-\frac{2x}{\breve{\ell}%
^{1/2}}+\frac{3x^{2}}{\breve{\ell}}+\frac{4x^{3}}{\breve{\ell}^{3/2}}+\cdots
,  \label{P3}
\end{equation}%
\begin{equation}
V\left( x\left( r\right) \right) =\frac{\breve{\ell}^{2}}{Q}\left[ V\left(
r_{\circ }\right) +V^{\prime }\left( r_{\circ }\right) \frac{r_{\circ }x}{%
\breve{\ell}^{1/2}}+V^{\prime \prime }\left( r_{\circ }\right) \frac{%
r_{\circ }^{2}x^{2}}{2\breve{\ell}}+V^{\prime \prime \prime }\left( r_{\circ
}\right) \frac{r_{\circ }^{3}x^{3}}{6\breve{\ell}^{3/2}}+\cdots .\right] .
\label{P4}
\end{equation}%
Moreover, let us expand the energy so that%
\begin{equation}
E_{k,\ell }=\frac{\breve{\ell}^{2}}{Q}\left[ E_{0}+\frac{E_{1}}{\breve{\ell}}%
+\frac{E_{2}}{\breve{\ell}^{2}}+\frac{E_{3}}{\breve{\ell}^{3}}+\cdots ,%
\right]   \label{P5}
\end{equation}%
where $Q$ is a constant parameter that scales the potential at large-$\ell
_{D}$ limit and is set equal to $\breve{\ell}^{2}$ at the end of the
calculation. Consequently, we may now re-write (\ref{P1}) as 
\begin{gather}
\left\{ -\partial _{x}^{2}+\left[ \breve{\ell}+\left( 2\beta +1\right) +%
\frac{\beta \left( \beta +1\right) }{\breve{\ell}}\right] \left( 1-\frac{2x}{%
\breve{\ell}^{1/2}}+\frac{3x^{2}}{\breve{\ell}}+\frac{4x^{3}}{\breve{\ell}%
^{3/2}}+\cdots \right) \right.   \notag \\
\left. +\frac{r_{\circ }^{2}\,\breve{\ell}}{Q}\left[ V\left( r_{\circ
}\right) +V^{\prime }\left( r_{\circ }\right) \frac{r_{\circ }x}{\breve{\ell}%
^{1/2}}+V^{\prime \prime }\left( r_{\circ }\right) \frac{r_{\circ }^{2}x^{2}%
}{2\breve{\ell}}+\cdots \right] \right\} \Phi _{k}\left( x\left( r\right)
\right) =\xi _{k}\Phi _{k}\left( x\left( r\right) \right)   \label{P6}
\end{gather}%
with 
\begin{equation}
\xi _{k}=\frac{r_{\circ }^{2}\,\breve{\ell}}{Q}\left[ E_{0}+\frac{E_{1}}{%
\breve{\ell}}+\frac{E_{2}}{\breve{\ell}^{2}}+\frac{E_{3}}{\breve{\ell}^{3}}%
+\cdots \right] .  \label{P7}
\end{equation}%
This equation is to be compared with that of the one-dimensional anharmonic
oscillator%
\begin{equation}
\left[ -\partial _{x}^{2}+\frac{1}{4}\omega ^{2}x^{2}+\varepsilon _{\circ
}+P\left( x\right) \right] \chi _{k}\left( x\right) =\lambda _{k}\chi
_{k}\left( x\right)   \label{P8}
\end{equation}%
with $P\left( x\right) $ is a perturbation given by%
\begin{eqnarray}
P\left( x\right)  &=&\breve{\ell}^{-1/2}\left[ \varepsilon _{1}x+\varepsilon
_{3}\,x^{3}\right] +\breve{\ell}^{-1}\left[ \varepsilon
_{2}x^{2}+\varepsilon _{4}\,x^{4}\right]   \notag \\
&&+\breve{\ell}^{-3/2}\left[ \delta _{1}x+\delta _{3}\,x^{3}+\delta
_{5}\,x^{5}\right] ++\breve{\ell}^{-2}\left[ \delta _{2}x^{2}+\delta
_{4}\,x^{4}+\delta _{6}\,x^{6}\right] .  \label{P9}
\end{eqnarray}%
It is obvious that equation (\ref{P8}) represents the one-dimensional Schr%
\"{o}dinger anharmonic oscillator that has been readily discussed by Imbo et
al. \cite{Imbo 1983,Imbo 1984} and by Mustafa and Barakat \cite{Mustafa 1998}
and reported to admit the eigenvalues%
\begin{equation}
\lambda _{k}=\varepsilon _{0}+\left( k+\frac{1}{2}\right) \,\omega +\frac{%
\alpha _{1}}{\breve{\ell}}+\frac{\alpha _{2}}{\breve{\ell}^{2}}.  \label{P10}
\end{equation}%
where $\alpha _{1}$ and $\alpha _{2}$ are given in the Appendix, and $\left(
k+\frac{1}{2}\right) \,\omega $ are the energies of the unperturbed
one-dimensional Schr\"{o}dinger oscillator with $k=0,1,2,\cdots $ that
denotes the number of nodes in the wave function (for $D=2,3$ it denotes the
radial quantum number or if you wish, the number of nodes in the wave
function). A comparison between (\ref{P6}) and (\ref{P8}) would imply that 
\begin{equation}
\lambda _{k}=\breve{\ell}\left( 1+\frac{r_{\circ }^{2}}{Q}V\left( r_{\circ
}\right) \right) +\left[ 2\beta +1+\left( k+\frac{1}{2}\right) \,\omega %
\right] +\frac{1}{\breve{\ell}}\left[ \beta \left( \beta +1\right) +\alpha
_{1}\right] +\frac{\alpha _{2}}{\breve{\ell}^{2}}+\cdots .  \label{P11}
\end{equation}%
We may now simply compare (\ref{P7}) with (\ref{P11}) to obtain%
\begin{equation}
E_{0}=\frac{Q}{r_{\circ }^{2}}+V\left( r_{\circ }\right) ,\;E_{1}=\frac{Q}{%
r_{\circ }^{2}}\left[ 2\beta +1+\left( k+\frac{1}{2}\right) \,\omega \right]
,\;E_{2}=\frac{Q}{r_{\circ }^{2}}\left[ \beta \left( \beta +1\right) +\alpha
_{1}\right] ,\;E_{3}=\frac{Q}{r_{\circ }^{2}}\alpha _{2},
\label{Es equations}
\end{equation}%
where $r_{\circ }$ is chosen to minimize the zeroth order term so that it
satisfies the relations%
\begin{equation}
\frac{dE_{0}}{dr_{\circ }}=0,\;\frac{d^{2}E_{0}}{dr_{\circ }^{2}}%
>0\Longrightarrow \left( \ell _{D}-\beta \right) =\sqrt{\frac{r_{\circ
}^{3}V^{\prime }\left( r_{\circ }\right) }{2}}.  \label{r0 eq}
\end{equation}%
and the shifting parameter $\beta $ \ is determined through the requirement
that the first order term vanishes. This would result in%
\begin{equation}
E_{1}=0\Longrightarrow \beta =-\frac{1}{2}\left( k+\frac{1}{2}\right)
\,\omega -\frac{1}{2},  \label{E1}
\end{equation}%
with%
\begin{equation}
\omega =2\sqrt{3+\frac{r_{\circ }V^{\prime \prime }\left( r_{\circ }\right) 
}{V^{\prime }\left( r_{\circ }\right) }.}  \label{omega eq}
\end{equation}%
One would therefore recast our anharmonic oscillator energies of (\ref{P5})
as%
\begin{equation}
E_{k}=E_{0}+\frac{1}{r_{\circ }^{2}}\left[ \beta \left( \beta +1\right)
+\alpha _{1}\right] +\frac{\alpha _{2}}{\breve{\ell}\,r_{\circ }^{2}}.
\label{PSLET total energy}
\end{equation}

Now we recall the corresponding Schr\"{o}dinger potential for our
KG-particles in the G\"{o}del SR-type cosmic string spacetime with a
Cornnell-type Lorentz scalar interaction of (\ref{Heun1}). This would
result, using (\ref{omega eq}),(\ref{E1}) and (\ref{r0 eq}), that%
\begin{equation}
\omega =2\,\sqrt{\frac{4\,\tilde{\Omega}^{2}r_{\circ }^{3}+3amr_{\circ
}^{2}-bm}{\tilde{\Omega}^{2}r_{\circ }^{3}+amr_{\circ }^{2}-bm}},\;\beta =-%
\frac{1}{2}\left( k+\frac{1}{2}\right) \,\omega -\frac{1}{2},\;\tilde{\Omega}%
=\Omega E  \label{omega1}
\end{equation}%
and%
\begin{equation}
\breve{\ell}=\left( \ell _{D}-\beta \right) =\sqrt{\tilde{\Omega}%
^{2}r_{\circ }^{4}+amr_{\circ }^{4}-bmr_{\circ }^{2}};\;\ell _{D}=\ell
+\left( D-3\right) /2.  \label{L2}
\end{equation}%
Various especial cases of such settings are given in the following
illustrative examples.

\section{Illustrative examples}

In this section, we use PSLET and consider the following illustrative
examples.

\subsection{KG-oscillator I as a topologically trivial  G\"{o}del-type
spacetime byproduct of (\protect\ref{KG-oscillator u})}

We hereby recall the KG-oscillator I as a trivial G\"{o}del-type spacetime
of section 2 described by (\ref{KG-oscillator u}) as%
\begin{equation}
\phi ^{\prime \prime }\left( u\right) -\tilde{\gamma}^{2}\,u^{2}\,\phi
\left( u\right) =-\lambda \,\phi \left( u\right) \,;\,\,\lambda =-\frac{%
\tilde{\beta}}{\gamma ^{2}}\,,\tilde{\gamma}^{2}=\frac{1}{\gamma ^{2}}%
\,.u=\gamma x+k_{y}.  \label{1D PSLET}
\end{equation}%
Obviously, the central repulsive/attractive core $\ell _{D}\left( \ell
_{D}+1\right) /$ $r^{2}$ in the $D$-dimensional Schr\"{o}dinger equation (%
\ref{D-dimensional Schrodinger}) must vanish to resemble the one dimensional
form of Schr\"{o}dinger equation (\ref{1D PSLET}). In this case, $%
D=1\Longrightarrow \ell _{D}\left( \ell _{D}+1\right) =\ell \left( \ell
-1\right) =0$ for $\ell =0$ and $\ell =1$ to denote even and odd parities $%
\left( -1\right) ^{\ell }$, respectively. Moreover, $u\equiv r\in \left(
-\infty ,\infty \right) $ and $u_{\circ }$ is determined by (\ref{r0 eq}) to
read $u_{\circ }=\sqrt{\breve{\ell}/\left\vert \tilde{\gamma}\right\vert }%
\in 
\mathbb{R}
;\;\breve{\ell}=\ell _{D}-\beta =\ell -1-\beta $. This would in turn imply
that $\omega =4$ by (\ref{omega eq}) and $\beta =-2k-3/2$ by (\ref{E1}).
Consequently, equations (\ref{Es equations}) would yield the zeroth-order
correction 
\begin{equation}
E_{0}=2\left\vert \tilde{\gamma}\right\vert \left( 2k+\ell +\frac{1}{2}%
\right) =\left\{ 
\begin{tabular}{ll}
$2\left\vert \tilde{\gamma}\right\vert \left( 2k+\frac{1}{2}\right) $ & ;
for even parity $\ell =0$ \medskip  \\ 
$2\left\vert \tilde{\gamma}\right\vert \left( 2k+\frac{3}{2}\right) $ & ;
for odd parity $\ell =1$%
\end{tabular}%
\right\} ,  \label{1D oscillator energy}
\end{equation}%
which results in a common form%
\begin{equation}
E_{0}=2\left\vert \tilde{\gamma}\right\vert \left( n+\frac{1}{2}\right)
;\;\,n=0,1,2,\cdots ,  \label{E0 HO}
\end{equation}%
with $E_{1},E_{2},$ and $E_{3}$ identically vanish. Therefore, the energies
associated with the shifted Schr\"{o}dinger oscillator are%
\begin{equation}
\lambda _{n}=2\left\vert \tilde{\gamma}\right\vert \left( n+\frac{1}{2}%
\right) ;\;n=0,1,2,\cdots .  \label{En Oscillator}
\end{equation}%
Consequently, the KG-oscillator I above admits the exact energies reported
in (\ref{Exact KG-HO-energies}) and goes through the same procedure
discussed in section 2-A.

\subsection{Quasi-free KG-oscillator in G\"{o}del SR-type cosmic string
spacetime (\protect\ref{Som metric})}

Next, we consider a KG-particle in G\"{o}del SR-type cosmic string spacetime
background described by (\ref{PDM-KG1}) with the assumption that $V\left(
r\right) =S\left( r\right) =0$ (no interaction potential is involved, hence
the notion "quasi-free KG-particle" is introduced). In this case one would
rewrite (\ref{PDM-KG1}) as 
\begin{equation}
\left\{ \partial _{r}^{2}-\frac{\left( \tilde{\ell}^{2}-1/4\right) }{r^{2}}-%
\tilde{\Omega}^{2}r^{2}+\tilde{\lambda}\right\} R\left( r\right) =0,
\label{KG eq3}
\end{equation}%
where%
\begin{equation}
\tilde{\lambda}=E^{2}-2\,\tilde{\ell}\,\Omega \,E-\left(
k_{z}^{2}+m^{2}\right) \text{; \ }\tilde{\ell}=\frac{\ell }{\alpha },\;%
\tilde{\Omega}^{2}=\Omega ^{2}E^{2}.  \label{KG eq3 parameters}
\end{equation}%
Which resembles the two-dimensional radial Schr\"{o}dinger-oscillator with
an irrational angular frequency $\tilde{\Omega}=\pm \left\vert \Omega
E\right\vert $ (now angular velocity). Under such settings, our PSLET
implies that for $D=2\Longrightarrow \ell _{D}=\left\vert \tilde{\ell}%
\right\vert -1/2$ (with $\tilde{\ell}$ denoting irrational magnetic quantum
number and $\tilde{\ell}$ replaces our $\ell $ of PSLET in (\ref%
{D-dimensional Schrodinger})), $%
\mathbb{R}
\ni r\in \left( 0,\infty \right) $ and $r_{\circ }$ is determined by (\ref%
{r0 eq}) to read $r_{\circ }=\sqrt{\breve{\ell}/\left\vert \tilde{\Omega}%
\right\vert }\in 
\mathbb{R}
$ with $\,\omega =4$ , $\beta =-2k-3/2$ by (\ref{omega1}), and $\breve{\ell}%
=\ell _{D}-\beta =2k+\left\vert \tilde{\ell}\right\vert +1$ by (\ref{L2}).
Consequently, equations (\ref{Es equations}) would yield the zeroth-order
correction%
\begin{equation}
E_{0}=2\left\vert \Omega E\right\vert \left( 2k+\left\vert \tilde{\ell}%
\right\vert +1\right) .  \label{quasi-free HO energies}
\end{equation}%
One should be aware that $r_{\circ }=\sqrt{\breve{\ell}/\left\vert \tilde{%
\Omega}\right\vert }\in 
\mathbb{R}
$ is manifestly a condition that secures the finiteness and square
integrability wave function, and $E_{1},E_{2},$ and $E_{3}$ identically
vanish.

Therefore, the energies associated with quasi-free KG-oscillator in G\"{o}%
del SR-type cosmic string spacetime are given by (\ref{KG eq3 parameters}) as%
\begin{equation}
E^{2}-2\,\tilde{\ell}\,\Omega \,E-\left( k_{z}^{2}+m^{2}\right) =2\left\vert
\Omega E\right\vert \left( 2k+\left\vert \tilde{\ell}\right\vert +1\right)
;\;\left\vert \Omega E\right\vert =\left\{ 
\begin{tabular}{l}
$+\Omega _{\pm }E_{\pm }\smallskip $ \\ 
$-\Omega _{\mp }E_{\pm }$%
\end{tabular}%
\right\} \geq 0.  \label{HO1}
\end{equation}%
This would allow us to obtain%
\begin{equation}
\begin{tabular}{ll}
$E_{\pm ,1}=\pm \left\vert \Omega \right\vert \,\grave{n}_{+}\pm \,\sqrt{%
\Omega ^{2}\,\grave{n}_{+}^{2}+k_{z}^{2}+m^{2}};\,\grave{n}%
_{+}=2k+\left\vert \tilde{\ell}\right\vert +\tilde{\ell}+1,$ & for $%
\left\vert \Omega E\right\vert =+\Omega _{\pm }E_{\pm }\smallskip $ \\ 
$E_{\pm ,2}=\pm \left\vert \Omega \right\vert \,\grave{n}_{-}\pm \,\sqrt{%
\Omega ^{2}\,\grave{n}_{-}^{2}+k_{z}^{2}+m^{2}};\,\grave{n}%
_{-}=2k+\left\vert \tilde{\ell}\right\vert -\tilde{\ell}+1,$ & for $%
\left\vert \Omega E\right\vert =-\Omega _{\mp }E_{\pm }$%
\end{tabular}%
.  \label{HO2}
\end{equation}%
However, we may rearrange such energies so that%
\begin{equation}
\begin{tabular}{ll}
$E_{k,\ell ,\pm }^{\left( +\right) }=\pm \left\vert \Omega \right\vert \,%
\grave{n}_{\pm }\pm \,\sqrt{\Omega ^{2}\,\grave{n}_{\pm }^{2}+k_{z}^{2}+m^{2}%
}\Rightarrow \left\{ 
\begin{tabular}{l}
$E_{k,\ell ,+}^{\left( +\right) }=+\left\vert \Omega \right\vert \,\grave{n}%
_{+}+\sqrt{\Omega ^{2}\,\grave{n}_{+}^{2}+k_{z}^{2}+m^{2}}$ \\ 
$E_{k,\ell ,-}^{\left( +\right) }=-\left\vert \Omega \right\vert \,\grave{n}%
_{-}-\,\sqrt{\Omega ^{2}\,\grave{n}_{-}^{2}+k_{z}^{2}+m^{2}}$%
\end{tabular}%
\right\} ;\medskip \,$ & for $\Omega =+\left\vert \Omega \right\vert $ \\ 
$E_{k,\ell ,\pm }^{\left( -\right) }=\pm \left\vert \Omega \right\vert \,%
\grave{n}_{\mp }\pm \,\sqrt{\Omega ^{2}\,\grave{n}_{\mp }^{2}+k_{z}^{2}+m^{2}%
}\Rightarrow \left\{ 
\begin{tabular}{l}
$E_{k,\ell ,+}^{\left( -\right) }=+\left\vert \Omega \right\vert \,\grave{n}%
_{-}+\sqrt{\Omega ^{2}\,\grave{n}_{-}^{2}+k_{z}^{2}+m^{2}}$ \\ 
$E_{k,\ell ,-}^{\left( -\right) }=-\left\vert \Omega \right\vert \,\grave{n}%
_{+}-\,\sqrt{\Omega ^{2}\,\grave{n}_{+}^{2}+k_{z}^{2}+m^{2}}$%
\end{tabular}%
\right\} ;$ & for $\Omega =-\left\vert \Omega \right\vert $%
\end{tabular}%
,  \label{HO3}
\end{equation}%
where in $E_{k,\ell ,\pm }^{\left( +\right) }$ and $E_{k,\ell ,\pm }^{\left(
-\right) }$ the superscripts $\left( +\right) $ and $\left( -\right) $
identify states with positive and negative vorticities, $\Omega =\Omega
_{+}=+\left\vert \Omega \right\vert $ and $\Omega =\Omega _{-}=-\left\vert
\Omega \right\vert $, respectively. Moreover, one should notice that $\,%
\grave{n}_{\pm }\left( \tilde{\ell}=-\left\vert \tilde{\ell}\right\vert
\right) =\grave{n}_{\mp }\left( \tilde{\ell}=\left\vert \tilde{\ell}%
\right\vert \right) $ and consequently $E_{k,\ell ,\pm }^{\left( +\right)
}\left( \tilde{\ell}=-\left\vert \tilde{\ell}\right\vert \right) =E_{k,\ell
,\pm }^{\left( -\right) }\left( \tilde{\ell}=\left\vert \tilde{\ell}%
\right\vert \right) $ indicating spacetime-type associated degeneracies
(STAD). Moreover, when this result is compared with that of Carvalho et al 
\cite{Carvalho 2014} (i.e., equation (14) in \cite{Carvalho 2014}) we
observe that not only the negative energies are missed but also the STADs
discussed above. Their results should, therefore, be generalized into those
reported in (\ref{HO3}). So should be the case with the results reported on
the linear confinement of a scalar particle in G\"{o}del-type space-time by
Vit\'{o}ria \cite{Vitoria 2018} (their Eq. (23)) and in the related comment
by Neto et al \cite{Neto 2020} (their Eq. (18)) for zero linear confinement.
That would also include the results reported by Ahmed \cite{Ahmed 2020},
where he also missed all nodeless (i.e., states with $n_{r}=0$) states.

\subsection{KG-Coulombic particle in G\"{o}del SR-type cosmic string
spacetime (\protect\ref{Som metric}) at zero vorticity, $\Omega =0.$}

Now, we consider a KG-particle in a G\"{o}del SR-type cosmic string
spacetime (\ref{Som metric}) at zero vorticity, $\Omega =0$, and subjected
to a Lorentz scalar potential $S\left( r\right) =\beta _{2}/r$ in (\ref%
{PDM-KG1}) to obtain 
\begin{equation}
\left\{ \partial _{r}^{2}-\frac{\left( \mathcal{L}^{2}-1/4\right) }{r^{2}}-%
\frac{\tilde{\beta}_{2}}{r}+\tilde{\lambda}\right\} R\left( r\right) =0,
\label{KG-Coulomb eq}
\end{equation}%
where%
\begin{equation}
\tilde{\lambda}=E^{2}-\left( k_{z}^{2}+m^{2}\right) \text{; \ }\tilde{\ell}=%
\frac{\ell }{\alpha },\;\mathcal{L}^{2}=\tilde{\ell}^{2}+\beta _{2}^{2},\;%
\tilde{\beta}_{2}=2m\beta _{2}.  \label{KG-Coulomb eq parameters}
\end{equation}%
In this case, our PSLET recipe above implies that $D=2\Longrightarrow \ell
_{D}=\left\vert \mathcal{L}\right\vert -1/2$ (with $\mathcal{L}$ now denotes
irrational magnetic quantum number and $\mathcal{L}$ replaces our $\ell $ of
PSLET in (\ref{D-dimensional Schrodinger})), $r\in \left( 0,\infty \right) $
and $r_{\circ }$ is determined by (\ref{L2}) to read $r_{\circ }=2\breve{\ell%
}^{2}/\tilde{\beta}_{2}$ with $\omega =2$, $\beta =-k-1$ by (\ref{omega1}),
and $\breve{\ell}=\ell _{D}-\beta =k+\left\vert \mathcal{L}\right\vert +1/2$%
. Consequently, equations (\ref{Es equations}) would yield%
\begin{equation}
E_{0}=-\frac{\tilde{\beta}_{2}^{2}}{4\left( k+\left\vert \mathcal{L}%
\right\vert +1/2\right) ^{2}}  \label{Exact KG-Coulomb energy}
\end{equation}%
with $E_{1},E_{2},$ and $E_{3}$ identically vanish. Therefore, the energies
associated with KG-Coulombic particle in G\"{o}del SR-type cosmic string
spacetime are given by 
\begin{equation}
E_{k,\ell }=\pm \sqrt{k_{z}^{2}+m^{2}-\frac{m^{2}\beta _{2}^{2}}{\left(
k+\left\vert \mathcal{L}\right\vert +1/2\right) ^{2}}}.
\label{Exact KG Coulomb energies}
\end{equation}%
One should notice that the result of the Heun polynomials solution for the
KG-Coulomb in  (\ref{Heun7}) for the case  $a=0=\Omega $ and $b=\beta _{2}$
does not match the exact solution (\ref{Exact KG Coulomb energies}), where
the effect of the Coulombic parameter $b=\beta _{2}$ have no contribution in
(\ref{Heun7}). 

\subsection{Massless KG-particles in G\"{o}del SR-type cosmic string
spacetime in a Cornell-type Lorentz scalar potential}

We hereby use the massless KG equation (\ref{massless KG}) with a
Cornell-type Lorentz scalar potential $S\left( r\right) =\beta _{1}r+\beta
_{2}/r$ to obtain 
\begin{equation}
\left\{ \partial _{r}^{2}-\frac{\left( \mathcal{L}^{2}-1/4\right) }{r^{2}}-%
\tilde{\Omega}^{2}r^{2}+\tilde{\lambda}\right\} R\left( r\right) =0,.
\label{massless KG1}
\end{equation}%
\begin{equation}
\tilde{\Omega}^{2}=\Omega ^{2}\,E^{2}+\beta _{1}^{2},\;\mathcal{L}^{2}=%
\tilde{\ell}^{2}+\beta _{2}^{2},\;\tilde{\ell}=\frac{\ell }{\alpha },\,\,%
\tilde{\lambda}=E^{2}-2\,\tilde{\ell}\,\Omega \,E-k_{z}^{2}-2\beta _{1}\beta
_{2}.  \label{massless KG1 parameters}
\end{equation}%
Then, our PSLET\ recipe implies that $D=2\Longrightarrow \ell
_{D}=\left\vert \mathcal{L}\right\vert -1/2$ (with $\mathcal{L}$ denoting
irrational magnetic quantum number and $\mathcal{L}$ replaces our $\ell $ of
PSLET in (\ref{D-dimensional Schrodinger})), where $r_{\circ }$ is
determined by (\ref{L2}) to read $r_{\circ }=\sqrt{\breve{\ell}/\left\vert 
\tilde{\Omega}\right\vert }\in 
\mathbb{R}
$ , $\,\omega =4$ and $\beta =-\left( 2k+3/2\right) $ by (\ref{omega1}) to
imply $\breve{\ell}=\ell _{D}-\beta =2k+\left\vert \mathcal{L}\right\vert +1$%
. Under such settings, one obtains%
\begin{equation}
E_{0}=2\,\left\vert \Omega E\right\vert \sqrt{1+\frac{\beta _{1}^{2}}{\Omega
^{2}\,E^{2}}}\,\left( 2k+\left\vert \mathcal{L}\right\vert +1\right)
\Longrightarrow \tilde{\lambda}=2\,\left\vert \Omega E\right\vert \sqrt{1+%
\frac{\beta _{1}^{2}}{\Omega ^{2}\,E^{2}}}\,\left( 2k+\left\vert \mathcal{L}%
\right\vert +1\right) ,  \label{massless 1}
\end{equation}%
where $E_{2}=0,$ $E_{3}=0$, and $\left\vert \tilde{\Omega}\right\vert
=\,\left\vert \Omega E\right\vert \sqrt{1+\beta _{1}^{2}/\Omega ^{2}\,E^{2}}%
\geq 0$ is used. Consequently, we get%
\begin{equation}
\tilde{\lambda}=E^{2}-2\tilde{\ell}\Omega E-\left( k_{z}^{2}+2\beta
_{1}\beta _{2}\right) =2\,\left\vert \Omega E\right\vert \sqrt{1+\frac{\beta
_{1}^{2}}{\Omega ^{2}\,E^{2}}}\,\left( 2k+\left\vert \mathcal{L}\right\vert
+1\right)   \label{massless 2}
\end{equation}%
to yield%
\begin{equation}
E_{\pm }^{2}-2\tilde{\ell}\,\Omega _{\pm }\,E_{\pm }-\left( k_{z}^{2}+2\beta
_{1}\beta _{2}\right) =2\Omega _{\pm }\,E_{\pm }\sqrt{1+\frac{\beta _{1}^{2}%
}{\Omega ^{2}\,E^{2}}}\,\left( 2k+\left\vert \mathcal{L}\right\vert
+1\right)   \label{massless 3}
\end{equation}%
for $\left\vert \Omega E\right\vert =\Omega _{\pm }\,E_{\pm }$, and%
\begin{equation}
E_{\pm }^{2}-2\tilde{\ell}\,\Omega _{\mp }\,E_{\pm }-\left( k_{z}^{2}+2\beta
_{1}\beta _{2}\right) =-2\Omega _{\mp }\,E_{\pm }\sqrt{1+\frac{\beta _{1}^{2}%
}{\Omega ^{2}\,E^{2}}}\,\left( 2k+\left\vert \mathcal{L}\right\vert
+1\right) .  \label{massless 4}
\end{equation}%
for $\left\vert \Omega E\right\vert =-\Omega _{\mp }\,E_{\pm }$ . However,
the two results above would eventually imply that%
\begin{equation}
E_{\pm }^{2}\mp 2\,\left\vert \Omega \right\vert \,E_{\pm }N_{\pm }-\left(
k_{z}^{2}+2\beta _{1}\beta _{2}\right) =0\Rightarrow E_{k,l,\pm }=\pm
\left\vert \Omega \right\vert N_{\pm }\pm \sqrt{\Omega ^{2}N_{\pm
}^{2}+k_{z}^{2}+2\beta _{1}\beta _{2}},\text{ }  \label{massless 5}
\end{equation}%
to read%
\begin{equation}
E_{k,l,+}^{\left( +\right) }=+\left\vert \Omega \right\vert N_{+}+\sqrt{%
\Omega ^{2}N_{+}^{2}+k_{z}^{2}+2\beta _{1}\beta _{2}}\text{ \ \ and \ }%
E_{k,l,-}^{\left( +\right) }=-\left\vert \Omega \right\vert N_{-}-\sqrt{%
\Omega ^{2}N_{-}^{2}+k_{z}^{2}+2\beta _{1}\beta _{2}}  \label{massless 6}
\end{equation}%
for $\Omega =\Omega _{+}=+\left\vert \Omega \right\vert $, and%
\begin{equation}
E_{\pm }^{2}\mp 2\,\left\vert \Omega \right\vert \,E_{\pm }N_{\mp }-\left(
k_{z}^{2}+2\beta _{1}\beta _{2}\right) =0\Rightarrow E_{k,l,\pm }=\pm
\left\vert \Omega \right\vert N_{\mp }\pm \sqrt{\Omega ^{2}N_{\mp
}^{2}+k_{z}^{2}+2\beta _{1}\beta _{2}},  \label{massless 7}
\end{equation}%
to read%
\begin{equation}
E_{k,l,+}^{\left( -\right) }=+\left\vert \Omega \right\vert N_{-}+\sqrt{%
\Omega ^{2}N_{-}^{2}+k_{z}^{2}+2\beta _{1}\beta _{2}}\text{ \ \ and \ }%
E_{k,l,-}^{\left( -\right) }=-\left\vert \Omega \right\vert N_{+}-\sqrt{%
\Omega ^{2}N_{+}^{2}+k_{z}^{2}+2\beta _{1}\beta _{2}},  \label{massless 8}
\end{equation}%
for $\Omega =\Omega _{-}=-\left\vert \Omega \right\vert $, where%
\begin{equation}
N_{\pm }=\sqrt{1+\frac{\beta _{1}^{2}}{\Omega ^{2}\,E^{2}}}\,\left(
2k+\left\vert \mathcal{L}\right\vert +1\right) \pm \tilde{\ell}.
\label{massless 9}
\end{equation}%
Again, one observes that $\,N_{\pm }\left( \tilde{\ell}=-\left\vert \tilde{%
\ell}\right\vert \right) =N_{\mp }\left( \tilde{\ell}=\left\vert \tilde{\ell}%
\right\vert \right) $ and consequently $E_{k,\ell ,\pm }^{\left( +\right)
}\left( \tilde{\ell}=-\left\vert \tilde{\ell}\right\vert \right) =E_{k,\ell
,\pm }^{\left( -\right) }\left( \tilde{\ell}=\left\vert \tilde{\ell}%
\right\vert \right) $ to represent spacetime-type associated degeneracies
(STAD).

\section{Concluding remarks}

In this paper, we started with KG-oscillators in a topologically trivial G%
\"{o}del-type spacetime background. A KG-oscillator I as a topologically
trivial G\"{o}del-type spacetime byproduct, and a KG-oscillator II as a
topologically trivial G\"{o}del-type spacetime plus momentum operator
deformation byproduct. We have used simple and straightforward textbook
procedures and reported, for the sake of scientific correctness, their exact
relativistic energies in (\ref{Sec2a1}), (\ref{Sec2a2}), and (\ref{Energy eq}%
). This would make the results reported by \cite{Ahmed 2018} on
KG-oscillators in topologically trivial G\"{o}del-type spacetime (\ref{Ahmed
metric1}) only partially correct and consequently should be generalized into
those reported in (\ref{Energy eq}). Next, we have discussed the drawbacks
associated with the power series (Frobinous method) expansion approach that
implied the biconfluent Heun functions/polynomials solution (\ref{Heun3}).
We have shown that for the effective interaction potential $V_{eff}\left(
r\right) =$ $\tilde{\Omega}^{2}r^{2}+2mar+2mb/r$ in the Schr\"{o}dinger-like
quantum model (\ref{Heun1}), the biconfluent Heun polynomial solution (\ref%
{Heun3}) cease to be able to address the especial case of KG-Coulombic
particle when $a=0=\Omega $. Consequently,\ the biconfluent Heun polynomial
solution (\ref{Heun3}) can only be classified as a conditionally exact
solution that may address the KG-oscillator (\ref{Heun5}), when $a=0=b$, but
tragically fails to address the KG-Coulomb case when $a=0=\Omega $. 

As an alternative method/technique, we have recollected, in short, the so
called pseudo perturbative shifted $\ell $ expansion technique PSLET (e.g., 
\cite{Imbo 1983,Imbo 1984,Mustafa 1993,Mustafa 1994,Mustafa 1991,Mustafa
1998,Mustafa 2000,Mustafa 2002,Znojil 2002}) to solve for the $D$%
-dimensional Schr\"{o}dinger-type equations (which is a common form the
relativistic wave equations collapse into, like (\ref{1D PSLET}), (\ref%
{Heun1}), etc). Some illustrative examples on the straightforward/simplistic
applicability of PSLET procedure are used. Amongst, a KG-oscillator as a
topologically trivial G\"{o}del-type spacetime byproduct of (\ref%
{KG-oscillator u}) that collapses into the one dimensional form of Schr\"{o}%
dinger equation, and three examples that resemble the radial two-dimensional
form of Schr\"{o}dinger equation: a Quasi-free KG-oscillator in G\"{o}del
SR-type cosmic string spacetime (\ref{Som metric}), a KG-Coulombic particle
in G\"{o}del SR-type cosmic string spacetime (\ref{Som metric}) at zero
vorticity $\Omega =0$, and a massless KG-particle in G\"{o}del SR-type
cosmic string spacetime in a Cornell-type Lorentz scalar potential. In all
of our illustrative examples above, the exact energies for the harmonic
oscillator (i.e., (\ref{1D PSLET}), (\ref{KG eq3}), and (\ref{massless KG1}%
)) like and Coulomb like (\ref{KG-Coulomb eq}) interactions are obtained
from the zeroth (leading) order correction, where all higher order
correction identically vanish. In connection with the effective potential $%
V_{eff}\left( r\right) =$ $\tilde{\Omega}^{2}r^{2}+2mar+2mb/r$\ in the Schr%
\"{o}dinger-like quantum model (\ref{Heun1}), nevertheless, one would need
to numerically solve for $\omega $ and $r_{\circ }$ in (\ref{omega1}) and (%
\ref{L2}) for each parametric value involved in the G\"{o}del SR-type cosmic
string spacetime metric (\ref{Som metric}). In this case, although tedious,
one would work out the corresponding eigenvalues for such a quantum
mechanical problem. The accuracy of PSLET is shown to be very satisfactory
and reliable for similar Schr\"{o}dinger-like models \cite{Mustafa
1998,Mustafa 2000,Mustafa 1993,Mustafa 1994,Mustafa 2002,Znojil 2002}. This
issue already lies far beyond the scope of the current proposal. 

Finally, we believe that the effects of gravitational fields introduced by
different spacetime structures can be clarified if only if the fundamental
models like the KG-oscillators and KG-Coulomb are comprehensively and
explicitly reported. To the best of our knowledge, the results of the
current methodical proposal have never been published elsewhere.

\section{Appendix: Explicit forms of $\protect\alpha _{1}$ and $\protect%
\alpha _{2}$}

In this section we give the explicit forms of $\alpha _{1}$ and $\alpha _{2}$
as given by \cite{Imbo 1984,Mustafa 1998}, 
\begin{equation}
\alpha _{1}=\left[ \left( 1+2k\right) e_{2}+3\left( 1+2k+2k^{2}\right) e_{4}%
\right] -\frac{1}{\omega }\left[ e_{1}^{2}+6\left( 1+2k\right)
e_{1}e_{3}+\left( 11+30k+30k^{2}\right) e_{3}^{2}\right] ,  \label{alpha 1}
\end{equation}%
\begin{eqnarray}
\alpha _{2} &=&\left[ \left( 1+2k\right) d_{2}+3\left( 1+2k+2k^{2}\right)
d_{4}+5\left( 3+8k+6k^{2}+4k^{3}\right) d_{6}\right]   \notag \\
&&-\frac{1}{\omega }\left[ \left( 1+2k\right) e_{2}^{2}+12\left(
1+2k+2k^{2}\right) e_{2}e_{4}+2e_{1}d_{1}+2\left(
21+59k+51k^{2}+34k^{3}\right) e_{4}^{2}\right.   \notag \\
&&\left. +6\left( 1+2k\right) e_{1}d_{3}+30\left( 1+2k+2k^{2}\right)
e_{1}d_{5}+6\left( 1+2k\right) e_{3}d_{1}+2\left( 11+30k+30k^{2}\right)
e_{3}d_{3}\right.   \notag \\
&&\left. +10\left( 13+40k+42k^{2}+28k^{3}\right) e_{3}d_{5}\right] +\frac{1}{%
\omega ^{2}}\left[ 4e_{1}^{2}e_{2}+36\left( 1+2k\right)
e_{1}e_{2}e_{3}\right.   \notag \\
&&\left. +8\left( 11+30k+30k^{2}\right) e_{2}e_{3}^{2}+24\left( 1+2k\right)
e_{1}^{2}e_{4}+8\left( 31+78k+78k^{2}\right) e_{1}e_{2}e_{4}\right.   \notag
\\
&&\left. +12\left( 57+189k+225k^{2}+150k^{3}\right) e_{3}^{2}e_{4}\right] -%
\frac{1}{\omega ^{3}}\left[ 8e_{1}^{3}e_{3}+108\left( 1+2k\right)
e_{1}^{2}e_{3}^{2}\right.   \notag \\
&&\left. +48\left( 11+30k+30k^{2}\right) e_{1}e_{3}^{3}+30\left(
31+109k+141k^{2}+94k^{3}\right) e_{3}^{4}\right] ,  \label{alpha2}
\end{eqnarray}%
with%
\begin{equation}
e_{j}=\varepsilon _{j}/\omega ^{j/2}\,;\;d_{i}=\delta _{i}/\omega
^{i/2}\;;\;j=1,2,3,4,\;i=1,2,3,4,5,6,  \label{e's & d's}
\end{equation}%
\begin{equation}
\begin{tabular}{llll}
$\varepsilon _{1}=-4\beta ,$ & $\varepsilon _{2}=6\beta ,\;$ & $\varepsilon
_{3}=-4+r_{\circ }^{5}V^{\prime \prime \prime }\left( r_{\circ }\right)
/6Q,\;$ & $\varepsilon _{4}=5+r_{\circ }^{6}V^{\prime \prime \prime \prime
}\left( r_{\circ }\right) /24Q,$%
\end{tabular}%
\end{equation}%
\begin{equation}
\begin{tabular}{ll}
$\delta _{1}=-2\left( \beta ^{2}-1/4\right) ,\;\delta _{2}=3\left( \beta
^{2}-1/4\right) ,\;$ & $\delta _{3}=-8\beta \medskip ,\;\delta _{4}=10\beta ,
$ \\ 
$\delta _{5}=-6+r_{\circ }^{7}V^{^{\prime \prime \prime \prime \prime
}}\left( r_{\circ }\right) /120Q,$ & $\delta _{6}=7+r_{\circ
}^{8}V^{^{\prime \prime \prime \prime \prime \prime }}\left( r_{\circ
}\right) /120Q.$%
\end{tabular}%
\end{equation}

\end{document}